\documentclass{article}

\usepackage{spconf}
\usepackage{graphicx}
\usepackage{booktabs}
\usepackage{amsmath}
\usepackage{hyperref}
\usepackage{microtype}
\usepackage{xcolor}
\usepackage[capitalize]{cleveref}
\usepackage{flushend}

\title{Absolute Quality Ratings of Speech Enhancement Systems by Listeners of Different Ages and Degrees of Hearing Loss}

\name{\shortstack{Matteo Torcoli$^1$, Chih-Wei Wu$^2$,
Andrea Esposito$^1$,  
Phillip A. Williams$^2$, 
Katrien Cambier$^{1,3}$,\\
William Wolcott$^2$,
Antonio Curci$^1$,
Nicholas S. Reed$^1$,
and Mark Laureyns$^{1,3}$}}
\address{$^1$Amplifon S.p.A., Milan, Italy \hspace{1cm} $^2$Netflix, Inc., Los Gatos, USA \\ $^3$Thomas More University College, Audiology Department, Antwerp, Belgium}

\begin{document}

\maketitle

\begin{abstract}
Speech Enhancement (SE) supports listening, particularly for older adults with age-related hearing loss. Yet, enhanced Speech Quality (SQ) is commonly evaluated by young normal-hearing listeners, and how their ratings translate to older adults remains under-explored.
We compared absolute SQ ratings
from $40$ younger normal-hearing listeners ($20$--$30$ years) and
67 older listeners ($60$--$95$ years) with diverse audiometric profiles, after screening.
Test materials comprised natural dialogues 
with realistic backgrounds. 
SQ differences between SE systems that were clear for younger listeners were smaller or inseparable in older groups, regardless of hearing status.
Hearing loss severity was associated with lower absolute ratings, but did not strongly modulate the contraction in separable SQ differences.
A small, audiometrically
mixed subgroup of older listeners showed younger-like rating patterns, suggesting
that peripheral audiology alone cannot explain the contraction.
\end{abstract}


\section{Introduction}

Older and younger listeners differ in how they perceive Speech Quality
(SQ)~\cite{arehart2010speechquality, kates2014hasqi, biberger2025audioquality}. 
Age-related hearing loss, affecting over $65\%$ of adults above $60$ years of age~\cite{who2021hearing}, partly explains this
pattern, but differences have also been observed with
minimal audiometric loss~\cite{banh2012age} and after matching for Pure Tone Average (PTA)~\cite{fullgrabe2015age},
possibly reflecting reduced time-frequency resolution or cognitive changes affecting sound perception. 
However, these studies gave Speech Enhancement (SE) limited or no attention. 
SE is a central component of technologies often used by older adults, e.g., hearing aids~\cite{moore2026hearing} and accessible-audio solutions~\cite{torcoli2021dialogplus}.
Yet, the SE literature commonly assesses SQ without specifically recruiting older listeners or listeners with hearing loss~\cite{li2026urgent, diehl2023restoring}. 
Widely used standards~\cite{ituP808,ituBS1534} explicitly require normal-hearing participants. 
The resulting data also drive objective SQ metrics~\cite{mittag2021nisqa, ragano2024scoreq}. 

Older listeners and listeners with hearing loss are more commonly involved in speech intelligibility evaluations~\cite{diehl2023restoring, akeroyd2023clarity}, 
but SQ is a related yet distinct attribute that can continue to improve after intelligibility approaches ceiling~\cite{arehart2022comparison, chen2022inqss}.
It thus remains under-explored how older adults perceive enhanced SQ, and how SE-system differences rated by younger normal-hearing listeners translate to them. 
Advances in SE driven by deep neural networks (DNNs)~\cite{li2026urgent,diehl2023restoring,zheng2023sixty}
make SQ beyond intelligibility increasingly important, especially for entertainment
and hearing aids, where audio should also be high-quality and enjoyable for all.

We address this gap by comparing SQ ratings of conditions including SE systems between younger normal-hearing listeners and listeners aged $60$ or older with diverse audiometric profiles, using varied, ecologically valid audio material.
\section{Listening Test}
\subsection{Listeners}

\begin{figure}[t]
  \centering
  \includegraphics[width=0.94\columnwidth]{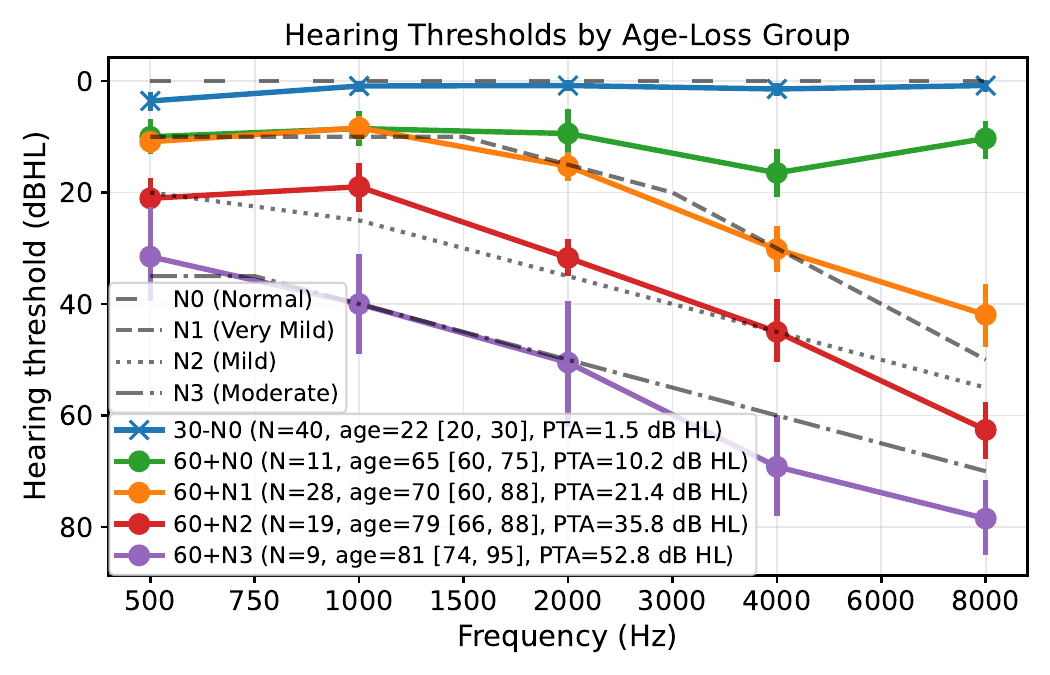}
  \caption{Reference and group-average audiograms with $95\%$ confidence intervals. Legends in this and subsequent figures report median age and age range.}
  \label{fig:pta_bisgaard}
\end{figure}

A listening test was conducted involving $133$ participants ($68$~female), none of whom used hearing aids: $42$ were aged $20$--$30$ years and $91$ were aged $60$--$95$ years and approximately evenly distributed across $60$--$69$, $70$--$79$, and $\geq 80$ years. Post-screening excluded $26$ participants with mean absolute PTA test-retest difference $> 5$~dB (\Cref{sec:method}), leaving $107$ participants ($52$~female).
Participants were grouped by age and the reference audiogram minimizing the root mean square error with the measured audiograms, as in~\cite{biberger2025audioquality}. The reference audiograms were Bisgaard N1--N3~\cite{bisgaard2010standard} with extrapolated $8$-kHz thresholds, plus N0 ($0$~dB~HL at all frequencies) representing normal hearing.
Following~\cite{who2021hearing}, 
grouping and PTA used the
better ear for bilateral loss and the worse ear for unilateral loss
(only one participant). 
Thus, we obtained five groups: relatively younger 30-N0 and older 60+N0 to N3 (\Cref{fig:pta_bisgaard}). 

Participants were recruited in Flanders, Belgium's Dutch-speaking region. On a self-reported scale
from 1 (not familiar at all) to 5 (extremely familiar), median English
familiarity was 5 for 30-N0 and 2--3 across the 60+ groups, whereas median
familiarity with Japanese and Korean was 1 for all groups.

\subsection{Audio Material and Conditions}
The main test used 20 source items (2.4--12.7~s) featuring 19 speakers
(8 female). 
Mostly drawn from a proprietary catalog of movies and TV shows, the material featured
expressive natural dialogues accompanied by environmental sounds, music, and sound
effects. The material comprised 50\% English-only excerpts, 25\% professional English dubs, and 25\% paired Korean or Japanese originals. 
These pairs were from animated productions and shared backgrounds, comparable emotional delivery, and dry studio speech recordings across languages, minimizing recording-related SQ differences.

The 20 unprocessed source items had a mean speech-to-background ratio of
$1.8\pm3.5$~dB, referred to as SNR, and were split into
two sets of 10, with participants randomly assigned to one set.
Each participant rated the resulting 90 main-test items (10 source items $\times$ 9
conditions) in randomized order, after 10 training items derived from two
additional source items.

The 9 conditions comprised 4 SE systems, 3 SE conditions under a so-called 
Hearing-Aid (HA) mode, referred to as SEHA, and 2 anchors.
All processed items were normalized to equal loudness following~\cite{ituBS1770}.
The SE systems included a Target SE
(TSE) condition that mixed clean speech with its background attenuated by 20~dB;
the other SE systems targeted the same
attenuation: 
SE1 and SE3 were the open-weight DNN-based systems
Open-Unmix~\cite{stoter2019openunmix} and
DeepFilterNet2~\cite{schroter2022deepfilternet2}, while SE2 used
an ideal phase-sensitive mask~\cite{erdogan2015phasesensitive} with coarse
frequency resolution (4-ms windows, 50\% overlap).

For the HA mode, the enhanced path was delayed by 10~ms,
amplified using the Clarity Challenge implementation~\cite{akeroyd2023clarity}
of NAL-R~\cite{byrne1986nal} for N3, and summed with the
unprocessed path.
The gain ranged from approximately 10~dB at 500~Hz to
24~dB at 6~kHz.
Subsequent loudness normalization retained the gain coloration but removed the overall level increase, providing a first, highly simplified approximation of the
coloration and comb-filtering effects that hearing aids may introduce. 
The lowest output SNR was 11~dB, comparable to the 
level reported as acceptable for older users of hearing aids in~\cite{schinkelbielefeld2023ceiling}. 

The anchors were: clean speech (the dry studio
recording without background audio) and reverb, generated using an extremely reverberant room impulse response ($T_{60}=7.6$~s) applied to TSE.
%
The SQ test used a sampling rate of 44.1~kHz; all source audio was originally sampled at 44.1~kHz or higher.

\subsection{Test Methodology}
\label{sec:method}

The test adopted an absolute quality paradigm following~\cite{ituBS1284}, similarly to prior studies~\cite{arehart2010speechquality,lundberg2020noise}.
Listeners were asked to rate SQ for each audio clip without direct
comparisons on a
continuous 0--100 quality scale, 
labeled with the qualitative descriptors \emph{bad}, \emph{poor},
\emph{fair}, \emph{good}, and \emph{excellent} to provide semantic anchors. The descriptors were translated into Dutch, following~\cite{zielinski2008biases}. Clips looped until rated or paused.

All listening sessions were conducted under quiet conditions using a portable setup with a calibrated iPad application: on campus for
younger participants and in their homes for older participants.
An experimenter supervised each session, available to assist.
Before the main SQ test, participants were given the possibility to adjust the playback volume to a comfortable level while listening to a dialogue clip that did not appear in the test. Levels were then fixed, and were relatively uniform across listeners, with 86\% within 4~dB of the calibrated default of 70~dB SPL.
%
%
The SQ rating was done using closed-back headphones (Beyerdynamic DT 770 Pro 32 ohm).

All participants underwent pure-tone audiometry at $0.5$, $1$, $2$, $4$, and $8$~kHz using calibrated audiometric headphones (Radioear dd65v2) both before and after SQ rating to validate the hearing-loss classification reliability. 

\section{Results}
\subsection{Statistical Analysis Framework}

Similarly to~\cite{arehart2022comparison,kates2025binaural}, SQ ratings were analyzed using Gaussian Linear Mixed-effects Models (LMMs) with listener and source item random intercepts. Age and audiogram were jointly encoded by the groups in \Cref{fig:pta_bisgaard}. For targeted comparisons, the same fixed-effects structure was fitted to restricted subsets. 
We also considered a full categorical by-listener condition-slope specification, but it yielded a singular fit and was excluded.
Still, listener-level variation is characterized in \Cref{sec:gain-model}.

Models and Type~III $F$-tests were fitted in R~4.5.2 using lme4/lmerTest with defaults: REML, Satterthwaite degrees of freedom, and the bobyqa optimizer. 
The primary fit converged without singularity. A logit-transformed analysis of the bounded scores yielded identical significance decisions ($\alpha=0.05$).
Adding the source-item split as a fixed effect showed no effect of the split itself ($p=0.899$) and left all significance decisions unchanged.

\begin{figure}[t]
  \centering
  \includegraphics[width=0.94\columnwidth]{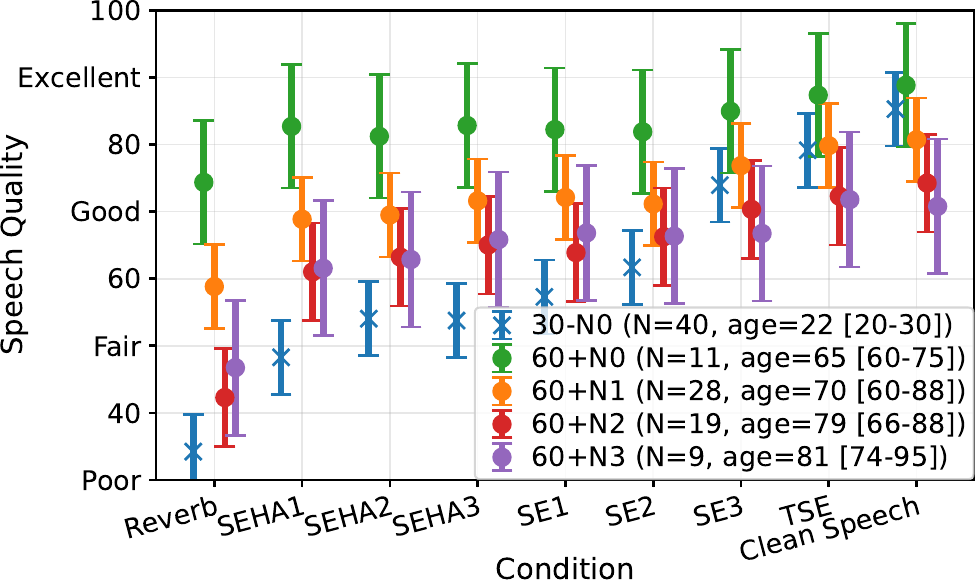}
  \caption{EMM condition ratings ($\pm95\%$ CI) by age-loss group.}
  \label{fig:ratings_overview}
\end{figure}

The results of the primary LMM analysis are in
Table~\ref{tab:omnibus}.
The significant factors are analyzed in the following subsections.
Figures~\ref{fig:ratings_overview} and~\ref{fig:clusters} show group $\times$ condition Estimated Marginal Means (EMMs) and $95\%$ Confidence Intervals~(CI) from the lmerTest/emmeans mixed-model fits.

\begin{table}[t]
  \centering
    \caption{Primary LMM analysis of the SQ ratings. Significant
  effects in bold ($p<0.05$). Effect size expressed in rating-scale points from EMMs: single contrast for Sex and Language; range of
  EMMs across levels for Condition and Age-Loss; range of per-level
  contrasts or spans for interactions.} 
  \label{tab:omnibus}
  \begin{footnotesize}
  \setlength{\tabcolsep}{2.5pt}
  \begin{tabular}{@{}lcccc c@{}}
    \toprule
    Effect & $F$ & df$_\mathrm{num}$ & df$_\mathrm{den}$ & Effect size & $p$ \\
    \midrule
    \textbf{Condition} & \textbf{181.2} & \textbf{8} & \textbf{9453.0} & \textbf{28.7} & \textbf{\textless0.001} \\
    \textbf{Age-Loss} & \textbf{7.0} & \textbf{4} & \textbf{102.2} & \textbf{22.1} & \textbf{\textless0.001} \\
    Language & 2.9 & 1 & 18.1 & 5.3 & $0.106$ \\
    Sex & 0.6 & 1 & 100.8 & 2.1 & $0.439$ \\
    \textbf{Condition $\times$ age-loss} & \textbf{26.8} & \textbf{32} & \textbf{9453.0} & \textbf{36.6} & \textbf{\textless0.001} \\
    \textbf{Language $\times$ age-loss} & \textbf{4.0} & \textbf{4} & \textbf{9458.7} & \textbf{6.2} & \textbf{0.003} \\
    Language $\times$ condition & 1.3 & 8 & 9453.0 & 4.0 & $0.243$ \\
    \bottomrule
  \end{tabular}
  \end{footnotesize}
\end{table}

\subsection{Overall Condition Ratings by Age-Loss Group}
\label{sec:age-loss}

\Cref{fig:ratings_overview} shows that group-average condition ratings differed in range and absolute level. Older listeners showed a clear \textit{contraction effect}, 
spanning a smaller range of the SQ scale, and
assigning smaller SQ differences between conditions than younger listeners. This occurred in all 60+ groups, including those not classified as having hearing loss.
Comparing 30-N0 and 60+N0, 
EMM for clean speech did not differ significantly between the two groups ($p = 0.41$), i.e., the two groups mostly differed in how strongly they penalized degradations.
Older listeners showed clear average SQ separation only
between extreme conditions, while 
SQ ratings differed markedly across almost all conditions for
younger listeners, who were consistent with
previous studies (our SE1 and SE3 correspond to Q1 and Q4 in~\cite{torcoli2024odaq}).

All Holm-corrected pairwise EMM
contrasts among the four core SE conditions (SE1, SE2, SE3, TSE) were significant for 30-N0, ranging from 21.9 points to 4.4 points.
However, the largest contrast in 30-N0 shrank to at most $8.4$ points in any 60+ group, and
its smallest contrast was not significant in any 60+ group ($-2.4$ to $1.0$ points, $p$ from $0.32$ to $1$).

Relative to the corresponding SE conditions, SEHA conditions were 
rated lower by the 30-N0 group, whereas the reduction was weak or inconsistent across the older groups.
SEHA differences were likewise compressed: in 30-N0, the largest pairwise contrast was $5.8$ points. 
This contraction was also observed in the older groups, but with larger uncertainty, motivating further study of how SQ
improvements reported under standard evaluation translate into assistive technologies. 

In addition to the contraction effect, \Cref{fig:ratings_overview} shows a distinct
\emph{lowering effect} within the older cohort: the condition-rating profiles
shifted downward with increasing
audiometric severity.
Condition-averaged EMM ratings were 10.1, 17.9, and 18.4 points lower than 60+N0 for 60+N1, 60+N2, and 60+N3, respectively ($p=0.046$, $p=0.001$, and $p=0.004$). 

\subsection{Modeling Per-Listener Contraction and Lowering}
\label{sec:gain-model}
To characterize contraction and lowering effects for individual listeners, we fitted the following per-listener LMM:
\begin{equation}
q_{cli}=a_l+b_l\mu_c+u_i+\varepsilon_{cli},
\label{eq:per_listener}
\end{equation}
where $q_{cli}$ is the SQ rating for condition $c$, listener $l$, and source item $i$;
$\mu_c$ is the centered empirical mean rating of condition $c$ in the 30-N0
group; $u_i$ is a source item random intercept; and $\varepsilon_{cli}$ is residual
error. 
The empirical 30-N0 condition means $\mu_c$ were computed once, centered across conditions, and used as fixed covariates without per-listener re-estimation.
Here, $a_l$ and $b_l$ are
the full listener-specific intercept and slope, including the corresponding
population fixed effects: $a_l=\beta_0+u_{0l}$ and
$b_l=\beta_1+u_{1l}$. Thus, the intercept $a_l$ 
summarizes the listener’s overall rating level, while the slope $b_l$ describes the
listener's scaling of condition differences relative to 30-N0.
The coefficients were obtained as best linear
unbiased predictors, with $(u_{0l},u_{1l})$ assumed to follow a
bivariate normal distribution with an estimated $2\times2$ covariance matrix.
%

\begin{figure}[t]
  \centering
  \includegraphics[width=0.95\columnwidth]{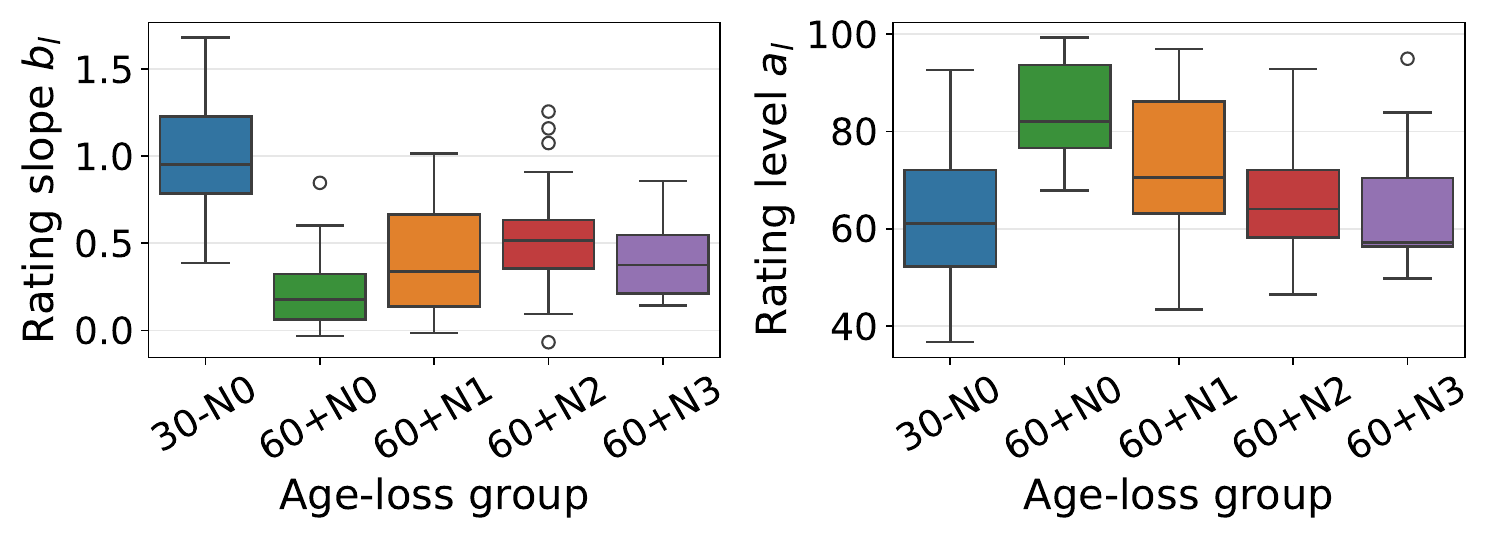}
  \caption{Per-listener rating slope $b_l$ and overall rating level $a_l$.}
  \label{fig:gain_model}
\end{figure}

\Cref{fig:gain_model} summarizes $b_l$ and $a_l$ by age-loss
group. All 60+
groups exhibited a substantially contracted slope. 
Within the 60+ cohort, $b_l$ was not significantly correlated with age
($\rho=0.075$, $p=0.55$) or English familiarity ($\rho=0.085$,
$p=0.49$), and only weakly correlated with PTA ($\rho=0.28$, $p=0.021$).
Restricting the considered data to only the four core SE systems, the weak PTA correlation was no longer significant ($p=0.26$),
suggesting that this association was driven by the more extreme conditions, and that PTA severity did not strongly modulate the contraction.

Moreover, the lowering effect was evident in the overall rating level
$a_l$, whose distribution shifted downward with increasing audiometric severity within the 60+ cohort. It correlated significantly with PTA ($\rho=-0.49$, $p<0.001$), but not with age ($\rho=-0.22$, $p=0.073$) or English familiarity ($\rho=0.06$, $p=0.61$), indicating that the lowering effect tracked audiometric severity within this cohort.

The strong contraction in 60+N0 may partly reflect a ceiling effect from its high ratings. However, all 60+ groups retained substantially smaller slopes than 30-N0 after logit transformation, suggesting that ceiling effects alone do not explain the contraction.

\subsection{Language in the Audio Clip}
Language had no significant main effect but a small, significant age-loss interaction (\Cref{tab:omnibus}). English dubs scored 4.5--7.2 points above paired originals, except in 60+N3 ($-0.5$). 
However, this group has a small sample size, warranting caution.


\section{Discussion}
Older listeners showed two complementary patterns: contracted SQ differences relative to the younger cohort, and, within the older cohort, lower absolute SQ ratings with increasing audiometric severity.
This lowering may partly reflect the lack of individualized hearing-loss compensation. Still, lower ratings have also been reported with \mbox{NAL-R} amplification~\cite{arehart2010speechquality}, suggesting that at least this linear amplification strategy may reduce but not eliminate the effect, possibly because it incompletely restores audibility or cannot reverse other consequences of hearing loss or age. 

While previous work associated hearing loss with lower ratings~\cite{arehart2010speechquality}, our finer grouping revealed this pattern consistently within the 60+ cohort, but not across generations: older groups with hearing loss rated some conditions higher than younger normal-hearing listeners, consistent with the contraction effect.

Earlier reports also attributed the contraction effect to hearing loss, as discussed in~\cite{arehart2010speechquality}, although that study did not observe it. 
We observed the contraction also for 60+N0, indicating that audiometric category alone cannot fully explain it.

\begin{figure}[t]
  \centering
  \includegraphics[width=0.95\columnwidth]{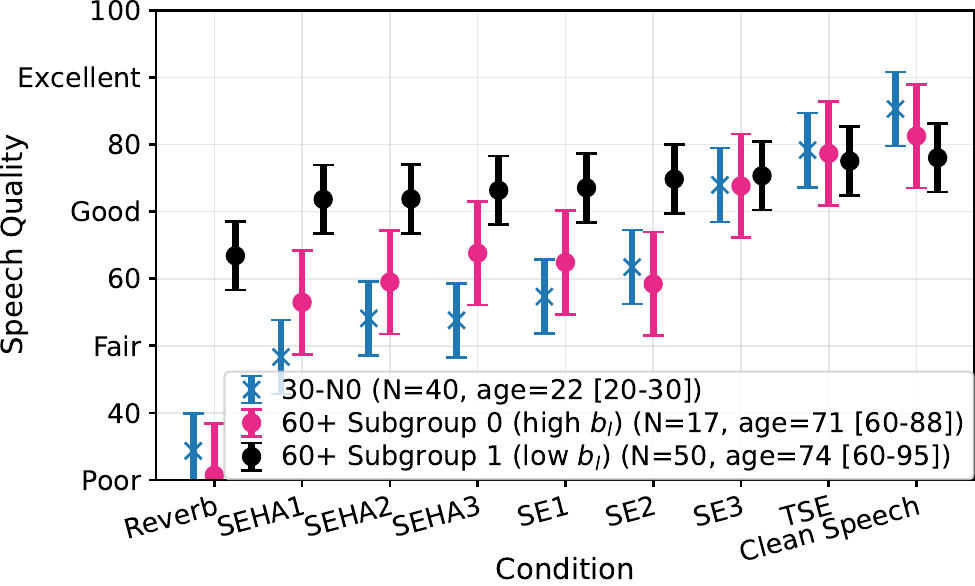}
  \caption{EMM condition ratings by slope-defined subgroups.}
  \label{fig:clusters}
\end{figure}

We divided the 60+ cohort into younger-like Subgroup~0 ($b_l\ge0.61$, the 15th percentile of \mbox{30-N0}) and Subgroup 1 (the remainder). \Cref{tab:cluster_profile} summarizes subgroups' characteristics and \Cref{fig:clusters} compares their ratings with 30-N0.
Holm-corrected pairwise EMM contrasts
among the four core SE conditions ranged from $3.2$ to $19.4$ points in
Subgroup~0 (5/6 contrasts significant), but from $1.3$ to $4.0$ points
in Subgroup~1 (1/6 significant).
%
Although exploratory because subgroups were defined from the same ratings, the presence of all 60+ age-loss classes in Subgroup~0 suggests that audiometry alone cannot explain the predominant contraction in Subgroup~1.

\begin{table}[t]
  \centering
  \caption{Subgroups composition within the 60+ cohort. 
  }
  \label{tab:cluster_profile}
  \begin{footnotesize}
  \setlength{\tabcolsep}{1.5pt}
  \newcommand{\stackhead}[2]{\begin{tabular}[t]{@{}c@{}}#1\\#2\end{tabular}}
  \begin{tabular}{@{}c|cccccc@{}}
    \toprule
    \stackhead{Sub-}{group} & \stackhead{N}{\phantom{group}} & \stackhead{Age}{median [range]} &
    \stackhead{PTA}{\phantom{group}} & \stackhead{Age-Loss (\%)}{N0/N1/N2/N3} &
    \stackhead{Sex (\%)}{F/M} & \stackhead{English}{familiarity} \\
    \midrule
    0 & 17 & 71 [60,88] & 29.3$\pm$11.5 & 6/53/29/12 & 35/65 & 2.71$\pm$1.36 \\
    1 & 50 & 74 [60,95] & 27.3$\pm$15.2 & 20/38/28/14 & 50/50 & 2.42$\pm$1.20 \\
    \bottomrule
  \end{tabular}
  \end{footnotesize}
\end{table}

To assess whether fatigue or declining concentration contributed to Subgroup~1's contraction, we compared listener slopes between trial halves. Changes were negligible and similar across subgroups ($-0.04$ vs.\ $-0.05$; Mann--Whitney $p=0.65$), providing no evidence of either explanation. Between-listener differences in overall concentration, engagement, or response bias remain possible.
Rating contraction has also been reported in crowdsourced experiments, motivating post-screening of listeners with contracted profiles~\cite{treffehn2026screening}. 
Our findings suggest that such screening could bias age-diverse panels if contraction reflects genuine cohort-specific perception.

Future work should address our study's limitations by testing native language, lower output SNRs, direct-comparison methods, and investigating the roles of differences in perceptual priorities, cultural background, subclinical hearing loss, central auditory processing, and cognition.
%
%


\section{Conclusion}
For conditions including SE systems with output SNR~$\ge11$~dB, absolute SQ differences evident in younger listeners were smaller or no longer significant in older listeners' group-average ratings, regardless of hearing status. 
A small older subgroup with mixed audiometric profiles showed younger-like ratings, suggesting explanations beyond audiometry. 
Greater audiometric severity within the older cohort was associated with lower ratings. 
These findings may motivate further research into age-representative SQ evaluation of SE systems beyond the current focus on younger normal-hearing listeners.

\section{Acknowledgment}

We thank Brechje Callebert, Jelle Essers, Laura Voet, Lise Deboel, Lotte Verhoeven, and Lunia Sleeckx for conducting the sessions, and Giorgio Romano, Carlo Baruffini, Diego Fogliata, and Gianna Lardaro from AmplifonX for their support.

\bibliographystyle{IEEEbib}
{\small
\bibliography{references}

@article{kates2025binaural,
  author  = {James M. Kates and Mathieu Lavandier and Ramesh Kumar Muralimanohar and Emily M. H. Lundberg and Kathryn H. Arehart},
  title   = {Binaural Speech Intelligibility for Combinations of Noise, Reverberation, and Hearing-Aid Signal Processing},
  journal = {PLoS ONE},
  volume  = {20},
  number  = {1},
  year    = {2025},
  doi     = {10.1371/journal.pone.0317266}
}

@inproceedings{torcoli2024odaq,
  author    = {Matteo Torcoli and Chih-Wei Wu and Sascha Dick and Phillip A. Williams and Mhd Modar Halimeh and William Wolcott and Emanu{\"e}l A. P. Habets},
  title     = {{ODAQ}: Open Dataset of Audio Quality},
  booktitle = {IEEE Int. Conf. Acoust. Speech Signal Process. (ICASSP)},
  pages     = {836--840},
  year      = {2024},
  doi       = {10.1109/ICASSP48485.2024.10447634}
}

@article{arehart2010speechquality,
  author  = {Kathryn H. Arehart and James M. Kates and Melinda C. Anderson},
  title   = {Effects of Noise, Nonlinear Processing, and Linear Filtering on Perceived Speech Quality},
  journal = {Ear Hear},
  volume  = {31},
  number  = {3},
  pages   = {420--436},
  year    = {2010},
  doi     = {10.1097/AUD.0b013e3181d3d4f3}
}

@article{lundberg2020noise,
  author  = {Emily M. H. Lundberg and Song Hui Chon and James M. Kates and Melinda C. Anderson and Kathryn H. Arehart},
  title   = {The Type of Noise Influences Quality Ratings for Noisy Speech in Hearing Aid Users},
  journal = {J. Speech Lang. Hear. Res.},
  volume  = {63},
  number  = {12},
  pages   = {4300--4313},
  year    = {2020},
  doi     = {10.1044/2020_JSLHR-20-00156}
}

@inproceedings{treffehn2026screening,
  author  = {Anika Treffehn and Andrea Eichenseer and Emily Kratsch and Nicola Pia},
  title   = {Screening Matters: A Comparative Study of Conventional and Crowdsourced Listening Tests},
  booktitle = {Interspeech},
  year    = {2026}
}

@article{diehl2023restoring,
  author  = {Peter Udo Diehl and Yosef Singer and Hannes Zilly and Uwe Sch{ö}nfeld and Paul Meyer-Rachner and Mark Berry and Henning Sprekeler and Elias Sprengel and Annett Pudszuhn and Veit M. Hofmann},
  title   = {Restoring Speech Intelligibility for Hearing Aid Users with Deep Learning},
  journal = {Sci. Rep.},
  volume  = {13},
  number  = {1},
  pages   = {2719},
  year    = {2023},
  doi     = {10.1038/s41598-023-29871-8}
}

@article{arehart2022comparison,
  author  = {Kathryn H. Arehart and Song Hui Chon and Emily M. Lundberg and Lewis O. {Harvey Jr.} and James M. Kates and Melinda C. Anderson and Varsha H. Rallapalli and Pamela E. Souza},
  title   = {A Comparison of Speech Intelligibility and Subjective Quality with Hearing-Aid Processing in Older Adults with Hearing Loss},
  journal = {Int. J. Audiol.},
  volume  = {61},
  number  = {1},
  pages   = {46--58},
  year    = {2022},
  doi     = {10.1080/14992027.2021.1900609}
}

@inproceedings{chen2022inqss,
  author    = {Yu-Wen Chen and Yu Tsao},
  title     = {{InQSS}: A Speech Intelligibility and Quality Assessment Model Using a Multi-Task Learning Network},
  booktitle = {Interspeech},
  pages     = {3088--3092},
  year      = {2022},
  doi       = {10.21437/Interspeech.2022-10153}
}

@book{who2021hearing,
  author    = {{World Health Organization}},
  title     = {World Report on Hearing},
  year      = {2021},
  isbn      = {978-92-4-002048-1}
}

@article{zheng2023sixty,
  author  = {Chengshi Zheng and Huiyong Zhang and Wenzhe Liu and Xiaoxue Luo and Andong Li and Xiaodong Li and Brian C. J. Moore},
  title   = {Sixty Years of Frequency-Domain Monaural Speech Enhancement: From Traditional to Deep Learning Methods},
  journal = {Trends Hear.},
  volume  = {27},
  pages   = {23312165231209913},
  year    = {2023},
  doi     = {10.1177/23312165231209913}
}

@inproceedings{torcoli2021dialogplus,
  author    = {Matteo Torcoli and Christian Simon and Jouni Paulus and Davide Straninger and Alfred Riedel and Volker Koch and Stefan Wirts and Daniela Rieger and Harald Fuchs and Christian Uhle and Stefan Meltzer and Adrian Murtaza},
  title     = {{Dialog+} in Broadcasting: First Field Tests Using Deep-Learning-Based Dialogue Enhancement},
  booktitle = {Int. Broadcasting Conv. (IBC)},
  year      = {2021}
}

@article{moore2026hearing,
  author  = {Brian C. J. Moore},
  title   = {Hearing Aids: What Works Well and What Can Be Improved},
  journal = {J. Assoc. Res. Otolaryngol.},
  volume  = {27},
  number  = {2},
  pages   = {123--135},
  year    = {2026},
  doi     = {10.1007/s10162-026-01031-5}
}

@inproceedings{li2026urgent,
  author    = {Chenda Li and Wei Wang and Marvin Sach and Wangyou Zhang and Kohei Saijo and Samuele Cornell and Yihui Fu and Zhaoheng Ni and Tim Fingscheidt and Shinji Watanabe and Yanmin Qian},
  title     = {{ICASSP} 2026 {URGENT} Speech Enhancement Challenge},
  booktitle = {IEEE Int. Conf. Acoust. Speech Signal Process. (ICASSP)},
  pages     = {21919--21921},
  year      = {2026},
  doi       = {10.1109/ICASSP55912.2026.11461686}
}

@techreport{ituP808,
  author      = {{Int. Telecom. Union}},
  title       = {Subjective Evaluation of Speech Quality with a Crowdsourcing Approach},
  institution = {ITU-T},
  type        = {Rec.},
  number      = {P.808},
  month       = jun,
  year        = {2021}
}

@techreport{ituBS1534,
  author      = {{Int. Telecom. Union}},
  title       = {Method for the Subjective Assessment of Intermediate Quality Level of Audio Systems},
  institution = {ITU-R},
  type        = {Rec.},
  number      = {BS.1534-3},
  month       = oct,
  year        = {2015}
}

@techreport{ituBS1284,
  author      = {{Int. Telecom. Union}},
  title       = {General Methods for the Subjective Assessment of Sound Quality},
  institution = {ITU-R},
  type        = {Rec.},
  number      = {BS.1284-2},
  month       = jan,
  year        = {2019}
}

@techreport{ituBS1770,
  author      = {{Int. Telecom. Union}},
  title       = {Algorithms to Measure Audio Programme Loudness and True-Peak Audio Level},
  institution = {ITU-R},
  type        = {Rec.},
  number      = {BS.1770-5},
  month       = nov,
  year        = {2023}
}

@inproceedings{mittag2021nisqa,
  author    = {Gabriel Mittag and Babak Naderi and Assmaa Chehadi and Sebastian M{ö}ller},
  title     = {{NISQA}: A Deep {CNN}-Self-Attention Model for Multidimensional Speech Quality Prediction with Crowdsourced Datasets},
  booktitle = {Interspeech},
  pages     = {2127--2131},
  year      = {2021},
  doi       = {10.21437/Interspeech.2021-299}
}

@inproceedings{ragano2024scoreq,
  author    = {Alessandro Ragano and Jan Skoglund and Andrew Hines},
  title     = {{SCOREQ}: Speech Quality Assessment with Contrastive Regression},
  booktitle = {Adv. Neural Inf. Process. Syst. (NeurIPS)},
  volume    = {37},
  pages     = {105702--105729},
  year      = {2024}
}

@article{banh2012age,
  author  = {Jessica Banh and Gurjit Singh and M. Kathleen Pichora-Fuller},
  title   = {Age Affects Responses on the Speech, Spatial, and Qualities of Hearing Scale ({SSQ}) by Adults with Minimal Audiometric Loss},
  journal = {J. Am. Acad. Audiol.},
  volume  = {23},
  number  = {2},
  pages   = {81--91},
  year    = {2012},
  doi     = {10.3766/jaaa.23.2.2}
}

@article{fullgrabe2015age,
  author  = {Christian F{\"u}llgrabe and Brian C. J. Moore and Michael A. Stone},
  title   = {Age-Group Differences in Speech Identification Despite Matched Audiometrically Normal Hearing: Contributions from Auditory Temporal Processing and Cognition},
  journal = {Front. Aging Neurosci.},
  volume  = {6},
  pages   = {347},
  month   = jan,
  year    = {2015},
  doi     = {10.3389/fnagi.2014.00347}
}

@article{biberger2025audioquality,
  author  = {Thomas Biberger and Stephan D. Ewert},
  title   = {Audio Quality Perception of Hearing-Impaired Listeners in Complex Acoustic Environments},
  journal = {Trends Hear.},
  volume  = {29},
  year    = {2025},
  doi     = {10.1177/23312165251374938}
}

@article{kates2014hasqi,
  author  = {James M. Kates and Kathryn H. Arehart},
  title   = {The Hearing-Aid Speech Quality Index ({HASQI}) Version 2},
  journal = {J. Audio Eng. Soc. },
  volume  = {62},
  number  = {3},
  pages   = {99--117},
  year    = {2014},
  doi     = {10.17743/jaes.2014.0006}
}

@article{zielinski2008biases,
  author  = {S{\l}awomir Zieli{\'n}ski and Francis Rumsey},
  title   = {On Some Biases Encountered in Modern Audio Quality Listening Tests -- A Review},
  journal = {J. Audio Eng. Soc. },
  volume  = {56},
  number  = {6},
  pages   = {427--451},
  year    = {2008}
}

@article{stoter2019openunmix,
  author  = {Fabian-Robert St{\"o}ter and Stefan Uhlich and Antoine Liutkus and Yuki Mitsufuji},
  title   = {Open-Unmix--A Reference Implementation for Music Source Separation},
  journal = {J. Open Source Software},
  volume  = {4},
  number  = {41},
  pages   = {1667},
  year    = {2019},
  doi     = {10.21105/joss.01667}
}

@inproceedings{schroter2022deepfilternet2,
  author    = {Hendrik Schr{\"o}ter and Alberto N. Escalante-B. and Tobias Rosenkranz and Andreas Maier},
  title     = {{DeepFilterNet2}: Towards Real-Time Speech Enhancement on Embedded Devices for Full-Band Audio},
  booktitle = {Int. W. on Ac. Sig. Enh. (IWAENC)},
  year      = {2022},
  doi       = {10.1109/IWAENC53105.2022.9914782}
}

@inproceedings{erdogan2015phasesensitive,
  author    = {Hakan Erdogan and John R. Hershey and Shinji Watanabe and Jonathan {Le Roux}},
  title     = {Phase-Sensitive and Recognition-Boosted Speech Separation Using Deep Recurrent Neural Networks},
  booktitle = {IEEE Int. Conf. Acoust. Speech Signal Process. (ICASSP)},
  pages     = {708--712},
  year      = {2015},
  doi       = {10.1109/ICASSP.2015.7178061},
}

@article{byrne1986nal,
  author  = {Denis Byrne and Harvey Dillon},
  title   = {The National Acoustic Laboratories' ({NAL}) New Procedure for Selecting the Gain and Frequency Response of a Hearing Aid},
  journal = {Ear Hear.},
  volume  = {7},
  number  = {4},
  pages   = {257--265},
  year    = {1986},
  doi     = {10.1097/00003446-198608000-00007}
}

@inproceedings{akeroyd2023clarity,
  author    = {Michael A. Akeroyd and Will Bailey and Jon Barker and Trevor J. Cox and John F. Culling and Simone Graetzer and Graham Naylor and Zuzanna Podwinska and Zehai Tu},
  title     = {The 2nd Clarity Enhancement Challenge for Hearing Aid Speech Intelligibility Enhancement: Overview and Outcomes},
  booktitle = {IEEE Int. Conf. Acoust. Speech Signal Process. (ICASSP)},
  year      = {2023},
  doi       = {10.1109/ICASSP49357.2023.10094918}
}

@article{schinkelbielefeld2023ceiling,
  author  = {Nadja Schinkel-Bielefeld and Jana Ritslev and Dina Lelic},
  title   = {Reasons for Ceiling Ratings in Real-Life Evaluations of Hearing Aids: The Relationship Between {SNR} and Hearing Aid Ratings},
  journal = {Front. Digit. Health},
  volume  = {5},
  pages   = {1134490},
  year    = {2023},
  doi     = {10.3389/fdgth.2023.1134490}
}

@article{bisgaard2010standard,
  author  = {Nikolai Bisgaard and Marcel S. M. G. Vlaming and Martin Dahlquist},
  title   = {Standard Audiograms for the {IEC} 60118-15 Measurement Procedure},
  journal = {Trends in Amplification},
  volume  = {14},
  number  = {2},
  pages   = {113--120},
  year    = {2010},
  doi     = {10.1177/1084713810379609}
}
}
\end{document}